\documentclass{appolb}
\usepackage{graphicx}

\begin{document}
\title{The method of an experimental determination of boundary conditions at a thin membrane for diffusion
\thanks{Presented at XXX Marian Smoluchowski Symposium on Statistical Physics}%
}
\author{Katarzyna D. Lewandowska
\address{Department of Radiological Informatics and Statistics,\\Medical University of
         Gda\'nsk,\\ ul. Tuwima 15, 80-210 Gda\'nsk, Poland}
\and
Tadeusz Koszto{\l}owicz
\address{Institute of Physics, Jan Kochanowski University,\\ul. \'Swi\c{e}tokrzyska 15, 25-406 Kielce, Poland}
}
\maketitle
\begin{abstract}
We present a method of deriving two boundary conditions at a thin membrane for diffusion from experimental data. This method can be really useful in complex membrane systems in which we do not know mechanisms of processes occurring within the membrane, since in such a situation the theoretical derivation of the boundary conditions seems to be impossible.  
\end{abstract}
  
\section{Introduction}

There are many systems in which substance is transported diffusively through a membrane. Such systems can be observed in life sciences as well as in engineering and they are widely discussed in the literature. We only mention here \cite{hr,l,h,kk,ck} and references cited therein.

The system under considerations consist of two parts divided by a thin membrane localized at $x=0$. All functions describing the process on the left-hand side of the membrane $(x<0)$ we denote with subscript $1$ and on the right-hand side of the membrane $(x>0)$ --- with subscript $2$. We assume that normal diffusion occurs in both parts of the system with diffusion coefficients $D_1$ in part $1$ and $D_1$ in part $2$. A real system is usually three--dimensional but we suppose that the considered system is homogeneous in the plane perpendicular to the $x$-axis and it can be treated as one-dimensional. Therefore, process of a substance transport can be described by the normal diffusion equations
\begin{eqnarray}
  \frac{\partial C_1(x,t)}{\partial t}&=&D_1\frac{\partial^2 C_1(x,t)}{\partial x^2}\;,\label{eq1a}\\
  \frac{\partial C_2(x,t)}{\partial t}&=&D_2\frac{\partial^2 C_2(x,t)}{\partial x^2}\;,\label{eq1b}
\end{eqnarray}
where $C_{1,2}(x,t)$ denotes a concntration profile in part $1$ and in part $2$, respectively. In order to solve these equations we need two boundary conditions at the membrane. The general form of the boundary conditions remains unknown, although there were many attempts to derive boundary conditions at the membrane, see, for example \cite{c,smol,koszt,koszt2001a,kdl,kwl} and the references cited therein. It should be mentioned here that one of the boundary conditions in a membrane system 
is usually assumed in the form that requires a continuity of a flux at the membrane $J_2(0^{+},t)=J_1(0^{-},t)$, where $J_{1,2}(x,t)=-D_{1,2}\partial C_{1,2}(x,t)/\partial x$, whereas the second boundary condition is most often chosen by an assumption. For example, in the case of a fully absorbing membrane there is $C(0,t)=0$ and for a fully reflecting wall we have $J(0,t)=0$ \cite{c,smol}. For a partially permeable membrane we can choose the second boundary condition as  $C_1(0^{-},t)/C_2(0^{+},t)=\kappa$, where $\kappa$ controls a membrane permeability \cite{koszt}, or in the form $J(0,t)=\kappa\left[C_1(0^{-},t)-C_2(0^{+},t)\right]$ \cite{koszt2001a}. We would like to point out that even in the case of the above mentioned boundary conditions occurring in relatively simple systems, there are ambiguous and nonequivalent. In more complicated systems such as, for instance, in a system with a membrane in which an absorption may occur, much more complex boundary conditions are expected. These boundary conditions are usually difficult to determine and they can take unexpected and astonishing forms. An example is the boundary condition in which the membrane permeability changes over time that reads $C_1(0^{-},t)=\lambda(t)C_2(0^{+},t)$, where $\lambda(t)=a+b\,{\rm exp}(wt)$, with $a$, $b$ and $w$ being constant \cite{kdl} or the boundary condition in the form of function quickly changing over time \cite{lk}. 

In this paper we present the method of boundary conditions determination at a thin membrane for diffusion from experimental data. Further considerations we perform within the Laplace transform domain ($\mathcal{L}\left\{f(t)\right\}\equiv\hat{f}(s)=\int_0^\infty{\rm e}^{-st}f(t)dt$), since it significantly simplify calculations. We consider the boundary conditions in general forms which read
\begin{eqnarray}
\hat{J}_2(0^+,s)&=&\hat{\Psi}(s)\hat{J}_1(0^-,s)\;,\label{eq3a}\\
\hat{C}_2(0^+,s)&=&\hat{\Phi}(s)\hat{C}_1(0^-,s)\;,\label{eq3b}
\end{eqnarray}
where $\hat{\Psi}(s)$ and $\hat{\Phi}(s)$ are functions that could be determined as follows.  Firstly, we choose some functions containing $\hat{\Psi}(s)$ and $\hat{\Phi}(s)$ that can easily be determined from experimental data. Then, we find theoretical formulae for them. Afterwords, we suggest that the same functions should be numerically determined from experimental data. Finally, the comparison of the theoretical version with the numerical version should allow the determination of $\hat{\Psi}(s)$ and $\hat{\Phi}(s)$ and, consequently, the boundary conditions (\ref{eq3a}) and (\ref{eq3b}).

\section{The method}
 
Normal diffusion equations (\ref{eq1a}) and (\ref{eq1b}) within the Lapalace transform domain take the forms
\begin{eqnarray}
  \hat{C}_1(x,s)-sC_1(x,0)&=&D_1\frac{\partial^2 \hat{C}_1(x,s)}{\partial x^2}\;,\label{eq3c}\\
  \hat{C}_2(x,s)-sC_2(x,0)&=&D_2\frac{\partial^2 \hat{C}_2(x,t)}{\partial x^2}\;,\label{eq3d}
\end{eqnarray}
where $C_{1,2}(x,0)$ denotes the initial concentration in part $1$ and $2$, respectively. We assume that the boundary conditions at the membrane have the forms (\ref{eq3a}) and (\ref{eq3b}).  We also suppose that particles move independently and do not clog the membrane, therefore the boundary conditions do not depend on an initial concentration. Thus, we choose the initial condition in a form that is convenient for experimental measurements which are often conducted by means of the laser interferometric method \cite{wadkg}. Namely, we assume that at the initial moment only part $1$ is filled with a diffusing substance, hence 
\begin{equation}
  \label{eq2}
  C_1(x,0)=C_0\;,\qquad C_2(x,0)=0\;.
\end{equation}
  
The Laplace transforms of solutions to Eqs.~(\ref{eq3c}) and (\ref{eq3d}) with the boundary conditions (\ref{eq3a}) and (\ref{eq3b}) and with the initial condition (\ref{eq2}) are 
\begin{eqnarray}
\hat{C}_1(x,s)&=&\frac{C_0}{s}\left[1-\frac{\sqrt{D_2}\hat{\Phi}(s)}{\sqrt{D_1}\hat{\Psi}(s)+\sqrt{D_2}\hat{\Phi}(s)}{\rm e}^{\sqrt{\frac{s}{D_1}}x}\right]\;,\label{eq4a}\\
\hat{C}_2(x,s)&=&\frac{C_0}{s}\frac{\sqrt{D_1}\hat{\Psi}(s)\hat{\Phi}(s)}{\sqrt{D_1}\hat{\Psi}(s)+\sqrt{D_2}\hat{\Phi}(s)}{\rm e}^{-\sqrt{\frac{s}{D_2}}x}\;.\label{eq4b}
\end{eqnarray}
The unknown functions $\hat{\Psi}(s)$ and $\hat{\Phi}(s)$ could be determined by comparison of our theoretical considerations with experimental data. This comparison would be much easier if we choose the following functions. Namely, on the left-hand side of the membrane ($x<0$) it is the time evolution of an amount of substance which leaves part $1$
\begin{equation}
  \label{eq4d}
W_1(t)=\int_{-\infty}^0\left[C_0-C_1(x,t)\right]dx\;,
\end{equation}
whereas on the right-hand side of the membrane ($x>0$) this function is the time evolution of an amount of substance which crosses the membrane
\begin{equation}
  \label{eq4c}
  W_2(t)=\int_0^\infty C_2(x,t)dx\;.
\end{equation}
The Laplace transform of (\ref{eq4d}) reads
\begin{equation}
\label{eq5}
\hat{W}_1(s)=\frac{C_0}{s^{3/2}}\frac{\sqrt{D_1D_2}\hat{\Phi}(s)}{\sqrt{D_1}\hat{\Psi}(s)+\sqrt{D_2}\hat{\Phi}(s)}\;,
\end{equation}
whereas the Laplace transform of (\ref{eq4c}) is
\begin{equation}
  \label{eq6}
\hat{W}_2(s)=\frac{C_0}{s^{3/2}}\frac{\sqrt{D_1D_2}\hat{\Phi}(s)\hat{\Psi}(s)}{\sqrt{D_1}\hat{\Psi}(s)+\sqrt{D_2}\hat{\Phi}(s)}\;.
\end{equation}
On the other hand, functions $\hat{W}_1(s)$ and $\hat{W}_2(s)$ could be obtained from experimental data by numerical calculating the Laplace transforms of $W_1(t)$ and $W_2(t)$ which, in turn, could be calculated from experimentally measured concentration profiles. Numerical calculations could be performed by means of, for example, the Gauss-Laguerre quadrature and the spline interpolation method \cite{sb}. A comparison of $\hat{W}_1(s)$ and $\hat{W}_2(s)$ obtained theoretically and numerically from experimental data would allow one to determine $\hat{\Psi}(s)$ and $\hat{\Phi}(s)$ and thereby to establish both boundary conditions at the membrane for diffusion.

A particular example of the application of the procedure presented above is the case considered in the paper \cite{kwl} in which we have presented the derivation of the second boundary condition from experimental data for the membrane system in which $D_1=D_2$. The first boundary condition was assumed in the form of (\ref{eq3a}) but for $\hat{\Psi}(s)=1$, whereas the second boundary condition was supposed as (\ref{eq3b}). The initial condition was chosen as (\ref{eq2}). Using the procedure specified earlier we obtained the following function $\hat{\Phi}(s)$ 
\begin{equation}
  \label{eq5a}
  \hat{\Phi}(s)=\frac{1}{\alpha+\beta\sqrt{s}}\;,
\end{equation}
where $\alpha$ and $\beta$ control the membrane permeability and the second boundary condition which took the form
\begin{equation}
  \label{eq2c}
  \alpha C_2(0^+,t)+\beta\frac{\partial^{1/2}}{\partial t^{1/2}}\;C_2(0^+,t)=C_1(0^-,t)\;,
\end{equation}
where $d^{1/2}f(t)/dt^{1/2}=(1/\sqrt{\pi})\left(d/dt\right)\int_0^tdt'f(t')/(t-t')^{1/2}$ denotes the Riemann--Liouville fractional derivative of the order $1/2$. The presence of a fractional derivative in the boundary condition is astonishing and shows that particles transfer through a thin membrane is a ``long-memory process'' even in the case of normal diffusion process.

\section{Final remarks}

We have presented the method of deriving two boundary conditions at a thin membrane for diffusion based on experimental data. We have proposed these boundary conditions in the general forms (\ref{eq3a}) and (\ref{eq3b}) but let us take note that many boundary conditions, some of which we have mentioned above, given in the terms of the Laplace transform can be expressed by Eqs.~(\ref{eq3a}) and (\ref{eq3b}). It should also be noticed that if $\hat{\Psi}(s)\neq1$ in (\ref{eq3a}) that means that the flux is not continuous at the membrane. In such a case, the form of the first boundary condition can be utterly astonishing. For example, a disturbance of the flux continuity may lead to surprising effects, such as the dependence of the flux on time. 

Many other unexpected or unusual effects resulting from, \textit{e.g.}, the lack of knowledge about processes occurring within a thin membrane, can affect the form of the boundary conditions and lead to a situation in which their theoretical form remain unknown. However, our method can give an answer to the question about the boundary conditions in all cases when we have experimental data. This method can be particularly useful when it is not known what processes take place within the membrane but an experiment is possible to conduct.

\section*{Acknowledgements}
This paper was partially supported by the Polish National Science Centre under grant No. 2014/13/D/ST2/03608.

\end{document}